\documentclass[sigconf]{acmart}

\AtBeginDocument{%
  \providecommand\BibTeX{{%
    \normalfont B\kern-0.5em{\scshape i\kern-0.25em b}\kern-0.8em\TeX}}}

\copyrightyear{2019}
\acmYear{2019}
\setcopyright{rightsretained}

\acmConference{ICSE 2020}{May 23-29, 2020}{Seoul, South Korea}

\usepackage{amsmath}
\usepackage{graphicx}
\usepackage{tikz}
\usetikzlibrary{shapes,backgrounds}
\usepackage{listings}
\begin{document}
\title{Sequence Model Design for Code Completion in the Modern IDE}

\author{Gareth Ari Aye}
\affiliation{%
  \institution{Google Inc., Columbia University}}
\email{aria@caa.columbia.edu}

\author{Gail E. Kaiser}
\affiliation{%
  \institution{Columbia University}}
\email{kaiser@cs.columbia.edu}

\begin{abstract}
Code completion plays a prominent role in modern integrated development environments (IDEs). Machine learning has become ubiquitous in analogous natural language writing and search software, surfacing more relevant autocompletions and search suggestions in fewer keystrokes. Prior research has reported training high-accuracy, deep neural networks for modeling source code, but little attention has been
given to the practical constraints imposed by interactive developer tools.

In particular, neural language models for source code modeling like the one described in \emph{Maybe Deep Neural Networks are the Best Choice for Modeling Source Code}\cite{DBLP:journals/corr/abs-1903-05734} are framed around code completion, but only report accuracy of next-token prediction. However, in order for a language model (LM) to work well within real-world code completion systems, it must also
\begin{itemize}
    \item always make suggestions that produce valid code that typechecks, to support code completion's role in correctness-checking,
    \item return instantaneous results to help programmers code more efficiently in fewer keystrokes, and
    \item be small enough to fit comfortably on disk and in memory on developer workstations, since virtually all modern IDEs run locally and support offline usage.
\end{itemize}

To meet these additional requirements, we propose a novel design for predicting top-$k$ next tokens that combines static analysis' ability to enumerate all valid keywords and in-scope identifiers with the ability of a language model to place a probability distribution over them. Our model mixes character-level input representation with token output to represent out-of-vocabulary (OOV) tokens meaningfully and minimize prediction latency. OOV tokens can be predicted through detection of local repetition common in software. This design achieves state-of-art accuracy in source code modeling and fits the constraints imposed by real-world code completion implementations in modern IDEs.
\end{abstract}



\maketitle

\section{Introduction}
Code completion is a tremendously popular tool for coding assistance, implemented across a wide range of programming languages and environments. In \emph{An Empirical Investigation of Code Completion Usage by Professional Software Developers}, Marasoiu et al. map out the diversity of use cases it fulfills for programmers, including correctness checking, typing assistance, and API search~\cite{muaruașoiuempirical}.  A study of programmers' behaviors within the Eclipse IDE found that autocomplete was used up to several times per minute~\cite{Murphy:2006:JSD:1159169.1159396}, as often as copy-paste!  Historically, completion suggestions have been based primarily on static analysis and, as a result, suffered from low relevance~\cite{Bruch:2009:LEI:1595696.1595728}. Applying the constraints imposed by a programming language's grammar and type system produces all valid suggestions but says nothing about which are likely.

\subsection{Language modeling}

In order to provide more relevant autocomplete results, researchers have looked to exploit the \textit{naturalness} of source code through language modeling~\cite{Hindle:2012:NS:2337223.2337322,Nguyen:2015:GSL:2818754.2818858}. Given a token sequence $S$ = $t_1t_2\cdot\cdot\cdot t_n$, a language model (LM) estimates the probability $p(S)$ as a product of conditional probabilities
\begin{center}
    $p(S) = \prod_{i=1}^n p(t_i | t_1,t_2,...,t_{i-1})$
\end{center}

State-of-art LMs for natural language are typically based on recurrent neural networks (RNNs) and have shown remarkable prediction accuracy in tasks ranging from machine translation to speech recognition~\cite{melis2018on}. Figure 3 depicts the basic RNN configuration: an input sequence $x_1, x_2, ..., x_n$ that the network learns to encode into hidden states $h_1, h_2, ..., h_n$ and decode as output. Each hidden state is computed as a combination of the previous hidden state and the next input sequence value.

In source code modeling, the input sequence $x_1, x_2, ..., x_n$ consists of vectors representing the previous $n$ tokens, abstract syntax tree (AST) nodes, characters, or even partial tokens. Commonly these inputs are represented by their integer index into an input vocabulary $V$ and the network will learn a dimensionality reduction $f(\textbf{x}) : {\rm I\!R}^{|V|} \rightarrow {\rm I\!R}^{m}$ for $m << |V|$ through an embedding layer. Whereas one-hot encoded vectors $\textbf{x},\ \textbf{x'} \in R^{|V|}$ have $\textbf{x} \perp \textbf{x'}$, the more compact vector space $R^m$ can capture semantic relationships among the transformed input vectors.

For the classification problem of selecting a high probability next word $x$ from a vocabulary $V$, the RNN is often connected to a softmax output and optimized to minimize cross-entropy loss. At a high-level, the softmax probability function $S(\textbf{x}) : {\rm I\!R}^{|V|} \rightarrow {\rm I\!R}^{|V|}$ produces an output vector $\textbf{y}$ so that $\sum_{i=1}^{|V|} y_i = 1$. This function allows the network to learn during training to decode the final RNN hidden state as probabilities of each word $x \in V$.

One of the most important choices in building an LM for source code is how this output vocabulary $V$ is constructed. As in the input sequence vocabulary, it could range from scanner tokens to individual characters. The model's vocabulary has implications for what can be predicted and how quickly predictions can be made. For example, a character-level model with $V = \{0, 1, ..., 255\}$ corresponding to ascii characters can output any arbitrary alphanumeric word, but requires numerous prediction requests. On the other hand, choosing $V$ to be the set of keywords for a given programming language makes it so only keywords and nothing else can be predicted, but whole tokens can be predicted in a single request.

\subsection{Modern IDE constraints}

Popular IDEs such as Visual Studio, IntelliJ, and Eclipse have in common that support for various programming languages is provided through a plugin architecture. This enables programmers to augment their IDE with additional language-specific functionality by downloading and installing extensions. These plugins provide interactive functionality to assist programmers writing software and include features like syntax highlighting, reporting of errors and warnings, and code completion.

\begin{figure}
    \centering
    \includegraphics[width=80mm]{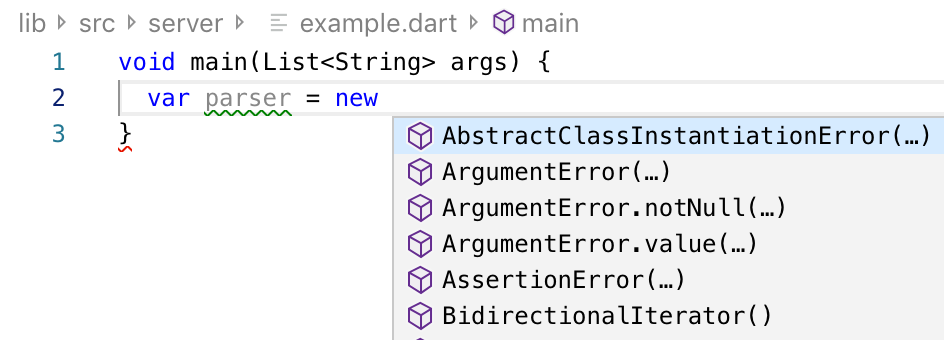}
    \caption{Completion suggestions before our ML ranking}
\end{figure}

\begin{figure}
    \centering
    \includegraphics[width=80mm]{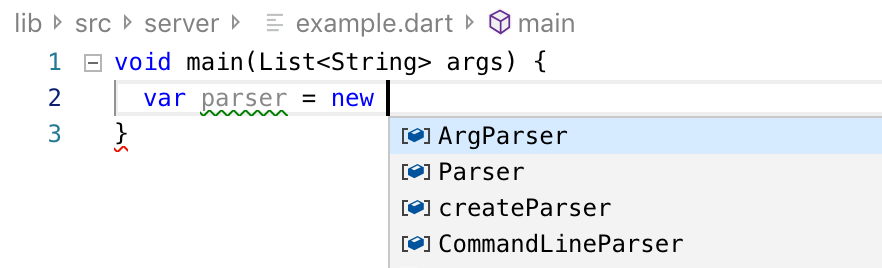}
    \caption{Completion suggestions after our ML ranking}
\end{figure}

Since one of the values that autocomplete provides is typing assistance, developers are interested in \emph{instantaneous} completion suggestions. The user experience literature states that completion results must arrive within 100ms of user action to be perceived as instantaneous~\cite{Miller:1968:RTM:1476589.1476628,Nielsen:1993:UE:529793}. This latency upper bound puts limits on LM size, the amount of processing that can be done before and after model predictions, and the number of predictions that can be made within an autocomplete request.

With regard to model size, deep neural networks for language modeling might contain hundreds of millions or even billions of parameters~\cite{DBLP:journals/corr/JozefowiczVSSW16}. Since each parameter represents a decimal value, an LM can quickly exceed the memory and disk capacity of a developer machine. Furthermore, a key finding from Jozefowicz et al. was that, given sufficient training data, the accuracy of RNN LMs improves with increasing size until a larger model cannot fit in GPU memory. In the context of IDE tools, accuracy improvements must be weighed against the resource costs of deploying and running a larger model on programmer workstations.

\begin{figure}
\hspace*{-0.5cm}
    \centering
    \includegraphics[scale=0.5]{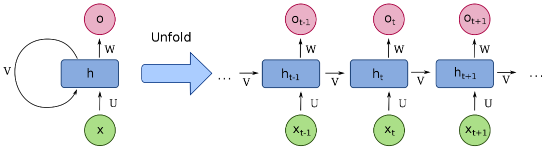}
    \caption{Unfolded recurrent neural network}
    \vspace{-10pt}
\end{figure}

Finally, programmers expect that accepting completion suggestions will, at the very least, produce programs that compile. As a matter of fact, Marasoiu et al. found in a study of Dart programmers' usage of code completion that developers often leverage autocomplete as a quick check of their code's correctness~\cite{muaruașoiuempirical}. In general, an LM cannot guarantee valid output. Penalizing invalid suggestions more heavily than valid but incorrect ones at training time by incorporating type analysis is an option, but would only decrease the likelihood of invalid suggestions. To ensure that suggestions are always valid, the model should be asked to rank already validated tokens or else any invalid suggestions it produces must be filtered out post-prediction.

\subsection{Summary of contributions}

\begin{itemize}
    \item This work details and evaluates a design for incorporating the predictive power of language modeling within existing IDE code completion systems.
    \item We discuss and compare prior work on neural source code modeling to address the open vocabulary problem.
    \item State-of-art accuracy is reported for source code modeling on a dynamic programming language.
    \item The largest of three corpora we studied is made available along with our source code~\cite{g_2019}.
\end{itemize}

This paper is organized to first give a brief history of the research that influenced our design. We then delve into the specifics of the design with a focus on modeling. We cover input representation, neural architecture, and training configurations, contrasting the details and performance of a character-input model with a token-input model. The design section concludes with a discussion of how model results can be synthesized with the list of keywords and in-scope identifiers produced by type analysis. We then characterize each of the datasets used for model training and evaluation, and report prediction accuracy in the context of comparable source code modeling results. An additional benchmark for prediction latency is reported to demonstrate the fitness of our approach for the IDE setting. Lastly, along with a concluding discussion, we consider threats to the validity of our work.

\section{Background}

The idea of leveraging machine learning to improve code completion suggestion ranking was proposed as early as 2009 in Bruch et al. \cite{Bruch:2009:LEI:1595696.1595728}, who considered the Eclipse IDE's unfortunate prioritization of all \textbf{java.lang.Object} methods when calling a method on any inheriting object. Hindle et al. connected this idea to the concept of natural language modeling and developed an $\emph{n}$-gram model for Java completion \cite{Hindle:2012:NS:2337223.2337322}. Raychev et al. and White et al. proposed replacing the $\emph{n}$-gram model for code completion with RNN \cite{Raychev:2014:CCS:2594291.2594321,White:2015:TDL:2820518.2820559}, which was reaffirmed as a superior tool in the 2019 paper \textit{Maybe Deep Neural Networks are the Best Choice for Modeling Source Code} \cite{DBLP:journals/corr/abs-1903-05734}.

\subsection{Pointer networks}
One of the major challenges researchers have encountered in applying LMs to software languages is that new words, especially identifier names, occur at a much faster rate than in natural language \cite{Allamanis:2013:MSC:2487085.2487127}, leading to a higher incidence of OOV tokens. One remediation for this issue is a pointer network \cite{vinyals2015pointer} where non-vocabulary tokens can be predicted by reference to an input sequence token. This strategy has been leveraged in multiple previous studies on source code modeling \cite{DBLP:journals/corr/BhoopchandRBR16,DBLP:journals/corr/abs-1711-09573}. There is a natural compatibility between pointer networks and source code because of the prevalence of locally repeated terms in software, but they do not address the problem of unknown token representation. The pointer network can only learn that a term appearing in one code location is likely to be repeated in another neighboring location.  

\subsection{Subword LM}
A more general idea from Karampatsis et al. is subword modeling \cite{DBLP:journals/corr/abs-1903-05734}, adapted from the work of Sennrich et al. on machine translation \cite{DBLP:journals/corr/SennrichHB15}. They represent source code through sequences of partial tokens and include special end-of-token symbols \lstinline{</t>} so that whole tokens can be constructed at prediction time. However, while this strategy solves the problem of OOV token representation, it introduces a new challenge at prediction time. Whereas a token-level LM only requires a single prediction, a subword model implies that a tree of partial, candidate suggestions must be searched to produce a list of high relevance results. Beam search \cite{6591953} is a natural fit for this task and terminates more quickly than a complete search, but subword predictions would nonetheless need to be extraordinarily fast in order to fit within the 100ms response window.

In order to get a sense for just how expensive it is to construct top-$k$ whole word predictions from a subword model, let's consider a very low entropy LM so that $p(x_i) = (\frac{1}{2})^i$ for $i \in [1, |V|]$. Let $m$ be the number of model subtokens forming a whole token. There are $|V|^m$ possible sequences $S$ of length $m$.

Suppose then in a realistic completion scenario that we want $k$ = 56 highest probability suggestions and that each token consists of $m$ = 3 model subtokens. Note that $m = 3$ is the smallest non-trivial choice and only consists of two partial tokens since the special token terminal symbol must be included. In order of decreasing probability, we need to find

\begin{itemize}
    \item 1 subtoken sequence $S_1$ with $p(S_1) = (\frac{1}{2})^3 = \frac{1}{8}$,
    \item $S_2, S_3, S_4$ with $p(S_i) = (\frac{1}{2})^2(\frac{1}{4}) = \frac{1}{16}$,
    \item $S_5, ..., S_{10}$ with $p(S_i) = (\frac{1}{2})^2(\frac{1}{8}) = (\frac{1}{2})(\frac{1}{4})^2 = \frac{1}{32}$,
    \item $S_{11}, ..., S_{20}$ with $p(S_i) = (\frac{1}{2})^2(\frac{1}{16}) = (\frac{1}{2})(\frac{1}{4})(\frac{1}{8}) = (\frac{1}{4})^3 = \frac{1}{64}$,
    \item $S_{21}, ..., S_{35}$ with $p(S_i) = (\frac{1}{2})^2(\frac{1}{32}) = (\frac{1}{2})(\frac{1}{4})(\frac{1}{16}) = (\frac{1}{2})(\frac{1}{8})^2 = (\frac{1}{4})^2(\frac{1}{8}) = \frac{1}{128}$, and
    \item $S_{36}, ..., S_{56}$ with $p(S_i) = (\frac{1}{2})^2(\frac{1}{64}) = (\frac{1}{2})(\frac{1}{4})(\frac{1}{32}) = (\frac{1}{2})(\frac{1}{8})(\frac{1}{16}) = (\frac{1}{4})^2(\frac{1}{16}) = (\frac{1}{4})(\frac{1}{8})^2 = \frac{1}{256}$.
\end{itemize}

With $i$ representing a selection of the $i$th highest probability subword, we then have

\begin{center}
    (1, 1, 1), (2, 1, 1), (1, 2, 1), (1, 1, 2), (3, 1, 1), (1, 3, 1), (1, 1, 3), (1, 2, 2), (2, 1, 2), (2, 2, 1), (4, 1, 1), (1, 4, 1), (1, 1, 4), (1, 2, 3), (1, 3, 2), (2, 1, 3), (2, 3, 1), (3, 1, 2), (3, 2, 1), (2, 2, 2), (1, 1, 5), (1, 5, 1), (5, 1, 1), (1, 2, 4), (1, 4, 2), (2, 1, 4), (2, 4, 1), (4, 1, 2), (4, 2, 1), (1, 3, 3), (3, 1, 3), (3, 3, 1), (2, 2, 3), (2, 3, 2), (3, 2, 2), (1, 1, 6), (1, 6, 1), (6, 1, 1), (1, 2, 5), (1, 5, 2), (2, 1, 5), (2, 5, 1), (5, 1, 2), (5, 2, 1), (1, 4, 3), (1, 3, 4), (4, 1, 3), (4, 3, 1), (3, 1, 4), (3, 4, 1), (2, 2, 4), (2, 4, 2), (4, 2, 2), (2, 3, 3), (3, 2, 3), (3, 3, 2)
\end{center}

We can make fewer predictions because of common prefixes. In fact, we only need in this example to make 28 subword predictions at (), (1), (2), (3), (4), (5), (6), (1, 1), (1, 2), (1, 3), (1, 4), (1, 5), (1, 6), (2, 1), (2, 2), (2, 3), (2, 4), (2, 5), (3, 1), (3, 2), (3, 3), (3, 4), (4, 1), (4, 2), (4, 3), (5, 1), (5, 2), and (6, 1). But even in this example of an artificially low-entropy LM with the smallest choice $m = 3$ of subwords per whole word, predictions would need an average speed of 3.57ms to meet the 100ms latency threshold. This is simply infeasible in a large RNN LM with $|V|$ = 10,000 like the best subtoken model reported by Karampatsis et al. \cite{DBLP:journals/corr/abs-1903-05734}.

\section{Design}

A hybrid strategy is to map input character sequences onto likely next tokens. Such a network combines the single-prediction speed of a token LM with the ability of a character-level model to meaningfully represent new words. The main drawback of this approach compared to a pure character-level LM like the one in Karampatsis et al. is that the network cannot predict novel words.

\subsection{Local repetition detection}

A solution inspired by the LM and pointer network combination of Bhoopchand et al. \cite{DBLP:journals/corr/BhoopchandRBR16} is to train an auxiliary network to predict whether the next token repeats one from the input sequence. Learning to predict repetition probability instead of the label's position within the input sequence is a better fit for our subtoken representation. This addition allows our network to assign probability to any token in the output vocabulary as well as any token appearing in the model's input sequence. It exploits the frequent occurrence of local repetition in source code.

Hellendoorn et al. found in their study of real-world code completions that intra-project completions are surprisingly common \cite{Hellendoorn:2019:CCF:3339505.3339625}; this component of our model is important as it allows project-specific tokens to be suggested. In the 
codebases we studied from GitHub, there is a 59.8\% probability that any keyword, identifier, or literal repeats one of the previous 100 tokens. At prediction time, given an estimate $\theta$ of the repetition probability, we can scale the probability mass assigned to tokens inside of the input sequence to $\theta$ and outside of the input sequence to $1 - \theta$. When probability mass is shifted onto input sequence tokens we can assign it specifically to OOV tokens that the fixed vocabulary models assign 0 probability.

\begin{table}[]
\hspace*{-0.5cm}
\begin{tabular}{|l|l|}
\hline
Feature                          & Motivation                                                                                                                                                 \\ \hline
Character-level model input      & \begin{tabular}[c]{@{}l@{}}Meaningfully represent new\\ words encountered at 
\\prediction time\end{tabular}                                                  \\ \hline
Token-level output               & \begin{tabular}[c]{@{}l@{}}Make predictions efficient by\\ only requiring a single 
\\prediction request\end{tabular}                                         \\ \hline
Repetition detection network     & \begin{tabular}[c]{@{}l@{}}Enable the model to assign\\ probability mass to OOV tokens,\\ leveraging frequent token 
\\repetition in source code\end{tabular} \\ \hline
Combination with static analysis & \begin{tabular}[c]{@{}l@{}}Ensure that predictions produce\\ code that typechecks\end{tabular}                                                             \\ \hline
\end{tabular}
\vspace{5pt}
\caption{Summary of design features}
\vspace{-20pt}
\end{table}

\subsection{Subtoken encoding}

One shortcoming of reading character rather than token input sequences is that the distance covered by \emph{n} time steps is much greater for \emph{n} tokens than for \emph{n} characters. RNN memory is limited in practice to tens or hundreds of time steps, and a typical line of code might contain up to 100 characters. This issue is addressed in \textit{Maybe Deep Neural Networks are the Best Choice for Modeling Source Code} \cite{DBLP:journals/corr/abs-1903-05734} through byte-pair encoding (BPE).

Rather than treat single characters as RNN input, in order to fit a longer sequence of preceding source code into fewer RNN time steps, we break identifier names into a short sequence of morphemes. A practice that is ubiquitous in source code is to concatenate terms like ``Directory'' and ``List'' into ``DirectoryList'' in order to construct compound names. There were two distinct styles present in the source code we studied, commonly referred to as \textbf{camelCase} and \textbf{snake\_case}. In the example encoding in Figure 4 this scheme allows our model to relate the lexemes ``ResourceProvider'' and ``FileResourceProvider''.

\begin{figure}
    \centering
    \begin{verbatim}
    [
        "class", "File", "Resource", "Provider",
        "implements", "Resource", "Provider", "{",
        "bool", "is", "_", "case", "_",
        "sensitive", ";",
    ]
    \end{verbatim}
    \caption{Example subtoken encoding}
    \vspace{-5pt}
\end{figure}

\subsection{RNN with GRU cells and projection layer}

Another challenge explored in the natural language realm is the computational cost of a neural network's softmax output layer when the number of classes is large. This difficulty arises from the need to compute each vocabulary word's activation, and significant attention has been given to strategies for replacing it with a faster scheme at training time including importance sampling \cite{bengio} and noise contrastive estimation \cite{gutmann2010noise}. However, while these techniques can speed up training, they have no effect on prediction speed.

One exception is the hierarchical softmax which replaces the network's flat output layer with a tree \cite{morin}. Instead of learning a single softmax function $S(\textbf{x}) : {\rm I\!R}^{|V|} \rightarrow {\rm I\!R}^{|V|}$, a softmax is learned and applied at every internal node to assign probability to each of its children. Then a prediction can be made without computing a probability for every single vocabulary word by following the highest probability path down the tree. Unfortunately this approach requires partitioning the output vocabulary. This conflicts with our need in code completion to return the top-\emph{k} maximum probability suggestions since we cannot ensure that they will share a leaf node across all predictions.

We employ Gated Recurrent Unit (GRU) cells \cite{DBLP:journals/corr/ChoMGBSB14} in our RNN LM as they achieved slightly superior performance and were faster to train than LSTM. Both LSTM and GRU cells address the problem of exploding and vanishing gradients in sequence models. These gated units allow the network to learn when to retain or forget hidden state information as the input sequence is processed \cite{Hochreiter:1997:LSM:1246443.1246450}. Figure 5 depicts a GRU cell where $x_t$ is a value from the input sequence and $h_{t-1}$ is the previous hidden state. These values are combined through reset gate vector $r_t$ and update gate vector $z_t$ to compute the recurrent activation $h_t$.

Additionally, a projection layer with linear activations is included between the network's hidden layer and softmax output as in Sak et al. \cite{DBLP:journals/corr/SakSB14} for lower latency. The projection layer significantly reduces the number of network parameters when the vocabulary is large.

\begin{figure}
    \hspace*{0.5cm}
    \centering
    \includegraphics[scale=0.5]{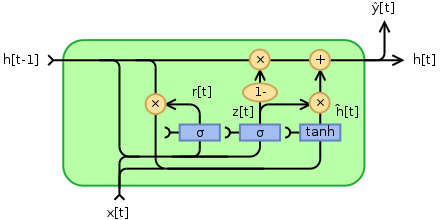}
    \caption{Gated recurrent unit}
\end{figure}

\subsection{Making predictions}

In order to make predictions from the token model as code completion is requested, a sequence of tokens $x_1, x_2, ..., x_{100}$ is created (inserting trailing padding if there are fewer than 100 previous tokens) starting at the cursor's file offset and walking toward the beginning of the file. A preprocessing step replaces $x_i \notin V$ with \textbf{<unk>}.

For the subtoken input model, each $x_i$ is encoded as a subsequence of partial tokens $m_1, m_2, ..., m_n$ by splitting on lower-to-uppercase transitions and underscore. OOV partial tokens encountered at prediction time are replaced with \textbf{<unk>}.

A parallel context then makes model predictions to estimate a probability mass function $p$ over the output vocabulary and a probability $\theta$ for repetition of some $x$ from the set $S$ of input sequence tokens. Then we can modify $p$ by scaling the output for elements $x$ so that $\sum_{x \in S} p(x)$ = $\theta$ and $\sum_{x \notin S} p(x)$ = $1 - \theta$. In the common case where probability mass is being shifted into $S$ we can use this opportunity to give non-zero probability to OOV tokens.

In the meantime, the main static analysis thread will work to enumerate all applicable keywords and identifiers in scope. Once the parallel prediction request returns, we can sort static analysis suggestions by decreasing model probability. In some cases static analysis may provide completion suggestions which are a concatenation of multiple tokens such as ``\lstinline{main()}'' which combines ``\lstinline{main}'', ``\lstinline{(}'', and ``\lstinline{)}''. The intention behind this behavior is to allow the user to complete the function's lexeme and add parentheses, indicating a function call, within a single completion acceptance. Since the completion model is only trained to predict single identifiers, keywords, and literals, we can match the outputs by checking that the concatenation begins with a token name from the model and that all subsequent characters do not belong to the identifier name alphabet (e.g. alphanumeric and underscore).

The tricky part is figuring out what to do with model-suggested tokens which haven't also been discovered as valid by type analysis. There are three possibilities to consider:

\begin{itemize}
    \item identifier declaration (e.g. new variable or function name),
    \item literal (e.g. string or integer), and
    \item invalid reference to an undefined or otherwise inapplicable token.
\end{itemize}

Our approach splits these model-only token suggestions based on whether they are literals. This can be determined with a simple regular expression to detect quotes or numeric lexemes. From there, the AST is checked to determine whether completion was requested in a declaration context (a code location where an identifier is being named). In a declaration context, only the non-literal suggestions are included. Otherwise the literal suggestions are added to the reranked suggestions from static analysis, unless a method is being invoked or an object attribute is being read, in which case all model-only suggestions are omitted.

\section{Datasets}
This work examines three large yet distinct corpora and considers the extent to which their token vocabularies overlap and autocompletion's performance on each of them.
One is the internal Dart repository of a large software company, referred to as {\em Internal} for double-blind review;
another is the set of open-source Dart code on GitHub; and the last consists of submissions to the "Flutter" mobile application development contest \cite{flutter}.

\begin{table}[]
\begin{tabular}{|l|l|l|}
\hline
Dataset         & Dart files & Unique training examples \\ \hline
Internal & 130K    & 27M      \\ \hline
GitHub          & 216K    & 19M      \\ \hline
Flutter Create  & 2K      & 1M       \\ \hline
\end{tabular}
\vspace{5pt}
\caption{Summary of corpora investigated}
\vspace{-20pt}
\end{table}

The Internal corpus is primarily comprised of web application code built with the AngularDart UI framework \cite{angulardart_dev_2016}. In contrast, the GitHub corpus consists of general-purpose frameworks and libraries that don't typically reference concepts specific to a single company.

Flutter is a cross-platform mobile application framework developed by Google and built with Dart. Flutter Create was an open programming contest that invited participants to create novel and high-quality mobile applications in 5kb of Dart or less \cite{fluttercreate}. It received nearly 1,000 submissions. While it is a smaller repository than the other two, it contains 
a diversity of domain-specific concepts. This aspect of the Flutter Create corpus was helpful in understanding model generalization and the extent to which OOV tokens hinder accuracy.

\subsection{Corpora statistics}

One important question is how similar each dataset is to the others. Overlap between their respective vocabularies is depicted in a Venn diagram in Figure 5. Looking at the simple intersections among the corpus vocabularies supports the argument that the problem of OOV tokens is indeed more severe in the programming language domain. Many of the vocabulary tokens are unique to their corpus -- 80\% for Internal,
71\% for GitHub, and 54\% for Flutter.

\begin{figure}
\begin{tikzpicture}
  \tikzset{venn circle/.style={draw,circle,minimum width=4cm}}
  \node [venn circle] (A) at (0,0) {$Internal\ (1,048,425)\ \ \ \ \ \ \ \ \ \ \ \ \ \ \ \ $};
  \node [venn circle] (B) at (60:2.67cm) {$Flutter\ (16,305)$};
  \node [venn circle] (C) at (0:2.67cm) {$\ \ \ \ \ \ \ \ \ \ \ \ \ \ \ \ \ \ GitHub\ (708,277)$};
  \node[left] at (barycentric cs:A=1/2,B=1/2 ) {$1,278$}; 
  \node[below] at (barycentric cs:A=1/2,C=1/2 ) {$254,236$};   
  \node[right] at (barycentric cs:B=1/2,C=1/2 ) {$1,767$};   
  \node[below] at (barycentric cs:A=1/3,B=1/3,C=1/3 ){$10,604$};
\end{tikzpicture}
\vspace{5pt}
\caption{Overlap among corpus' token vocabularies}
\vspace{-10pt}
\end{figure}
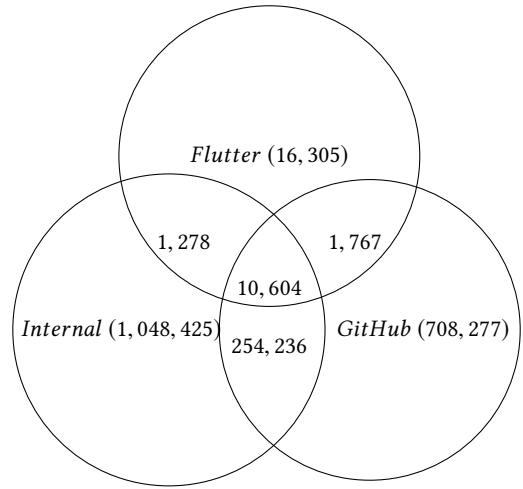

However, even though the 267,885 keywords, identifiers, and literals which appear in multiple corpora account for only 13.1\% of the union, they hold 90.6\% of the overall probability mass when looking at frequency of occurrence. This shows that, although identifier vocabulary size is theoretically unbounded and grows more quickly than in natural language vocabularies, a completion model with output limited to a fixed set of common tokens and local references can be viable.

\section{Methodology}

A token scanner is applied to transform each source file into a list of tokens. Each corpus-specific RNN model selected the most frequently occurring 100k tokens for the output vocabulary. These top tokens hold 0.8742 of the overall probability mass in the GitHub corpus and 0.8424 in the 
Internal corpus. The Flutter Create codebase contains fewer than 100k unique tokens.

Rare tokens are replaced by a special \textbf{<unk>} symbol in order to facilitate training. In order to construct training examples for next-token prediction, each keyword, identifier, or literal was treated as a label for the previous 100 inputs -- tokens for one model and partial tokens for the other. Duplicate examples (padded sequences of 101 matching items) within each of the corpora were removed. This preprocessing step is important for source code models because of the prevalence of code duplication \cite{DBLP:journals/corr/abs-1812-06469} and ensures that examples are distinct across our training and evaluation datasets. The scanned token sequences in each corpus were split into train and test sets. Two RNNs, respectively taking token and character input sequences, were trained on each of the three corpora. The common network architectures and training strategies are described below.

\subsection{Pure token-level RNN}

\begin{itemize}
    \item \textbf{Neural architecture.} Input layer embeds tokens as vectors with length 512. A single RNN layer containing 1024 GRU units is unrolled for 100 time steps. The hidden layer is projected onto a dense layer with 512 linear activation units to speed up training and prediction, as suggested in \textit{Long Short-Term Memory Based Recurrent Neural Network Architecture for Large Vocabulary Speech Recognition} \cite{DBLP:journals/corr/SakSB14}. A softmax over the corpus' most common 100k tokens computes a probability for each token.
    \item \textbf{Regularization.} The token network consists of over 100M parameters and significant regularization is necessary to prevent overfitting. Batch normalization \cite{DBLP:journals/corr/IoffeS15} is applied before and after the recurrent layer to normalize activations. Additionally, 0.5 dropout \cite{DBLP:journals/corr/abs-1207-0580} is applied before the final softmax layer following guidance from \textit{Understanding the Disharmony between Dropout and Batch Normalization by Variance Shift} \cite{DBLP:journals/corr/abs-1801-05134}. Dropout forces the network to learn robust features as opposed to simply memorizing the training dataset by randomly zeroing layer inputs.
    \item \textbf{Training strategy.} The network is optimized using stochastic gradient descent (SGD) with Nesterov momentum \cite{nesterov} of 0.9 and a batch size of 32 over 10 epochs. Gradient norms are clipped to 1.0 as proposed in Pascanu et al. \cite{DBLP:journals/corr/abs-1211-5063} to address the issue of numerical stability.
\end{itemize}

As in other RNN LM studies, training is very time consuming, and determining the optimal configuration of hyperparameter values is infeasible. Instead, we tested a few common values for each hyperparameter while fixing the others and selected the best performing value.

After training, the TensorFlow Lite (``tflite'') converter is applied to reduce network sizes from nearly 1Gb to around 100Mb to better fit on programmers' machines. The converter leverages post-training quantization to lower the amount of data stored for each network parameter with minimal impact on accuracy \cite{DBLP:journals/corr/abs-1712-05877}.

\subsection{Hybrid subtoken input, token output RNN}

Input training tokens are transformed into one or more partial tokens by splitting compound identifier names. Other details are the same as in the token-level model.

\subsection{Local repetition detection}

In both configurations, whether token or partial token input, an auxiliary network was trained to estimate the probability that the next token repeats one of the previous 100. It uses the same input sequence and internal representation as the LM, but the hidden layer is decoded using a single sigmoid function $\sigma(x) = \frac{1}{1 + e^{-x}}$ rather than a softmax output since an estimate $\theta$ of the repetition probability is desired rather than the probabilities $p(x)$ $\forall x \in V$. Training and prediction speed are significantly faster than in the LM, and the overall network is an order of magnitude smaller.

\section{Evaluation}

In order to assess model quality, top-1 and top-5 accuracy were measured on the two model variants using each corpus' held-out test data. Top-$k$ accuracy measures the percent of test examples in which one of the model's top $k$ predictions matches the ground-truth label.

In addition, since low latency is so important for user experience, a benchmark was created in which 10,000 prediction requests are made serially and the duration for each is recorded. The benchmark's input sequences were sampled directly from the corpora to make the performance simulation more realistic.

\section{Results}

Top-1 and top-5 accuracy are broken down across the three corpora and two architectures in Table 3 and 4. In our experiments, the partial token model outperforms the token model and accuracy increases as the corpus size grows. These results mirror the findings in Jozefowicz et al. where the best large scale natural language models leverage character-level embedding. The predictive ability of static analysis for code completion was measured on 
Internal for baseline comparison and shows the significant improvement achieved through language modeling.

\begin{table}[]
\begin{tabular}{|l|l|l|l|}
\hline
        & Token model & Subtoken model & Type analysis \\ \hline
Internal  & 0.6782 & 0.7139 & 0.0547 \\ \hline
GitHub  & 0.6687 & 0.6952 & \\ \hline
Flutter & 0.4848 & 0.4925 & \\ \hline
\end{tabular}
\vspace{5pt}
\caption{Top-1 accuracy (keywords, identifiers, and literals)}
\vspace{-10pt}
\end{table}

\begin{table}[]
\begin{tabular}{|l|l|l|l|}
\hline
        & Token model & Subtoken model & Type analysis \\ \hline
Internal  & 0.8465 & 0.886 & 0.1250 \\ \hline
GitHub  & 0.8276 & 0.8596 & \\ \hline
Flutter & 0.6744 & 0.7043 & \\ \hline
\end{tabular}
\vspace{5pt}
\caption{Top-5 accuracy (keywords, identifiers, and literals)}
\vspace{-10pt}
\end{table}


\section{Comparison to state of the art}

\emph{Learning Python Code Suggestion with a Sparse Pointer Network} \cite{DBLP:journals/corr/BhoopchandRBR16} and \emph{Code Completion with Neural Attention and Pointer Networks} \cite{DBLP:journals/corr/abs-1711-09573} are two recent studies that demonstrated state-of-art accuracy predicting next tokens in dynamically typed programming languages. They each report top-1 accuracy on Python codebases. The former also reports top-5 accuracy and the latter includes results from applying the \emph{PHOG: Probabilistic Model for Code} \cite{bielik2016phog} model to their Python corpus. Our partial token LM exceeded both results when trained and evaluated on the open-source GitHub Dart corpus.

\begin{table}[]
\begin{tabular}{|l|l|l|l|}
\hline
Model & Acc & Acc@5 \\ \hline
Our best & 0.7139 & 0.886 \\ \hline
Li et al. (2018) & 0.701 & -- \\ \hline
Bhoopchand et al. (2017)  & 0.6321 & 0.8262 \\ \hline
Raychev et al. (2016) & 0.692 & -- \\ \hline
\end{tabular}
\vspace{5pt}
\caption{Comparison to state-of-art source code modeling}
\vspace{-10pt}
\end{table}

Unfortunately the corpora and kinds of tokens which are included in source code model accuracy measurements vary greatly between studies. While some models might predict all tokens including punctuation, literals, and declaration names, others might not include any of these token types. This lack of consensus makes direct comparison a challenge. For instance, Raychev et al. \cite{Raychev:2014:CCS:2594291.2594321} report 0.9 top-3 accuracy but only on 20 test samples all of which are method invocations. On the other hand, Karampatsis et al. predict all token types in a large random sample of open-source Java corpora \cite{DBLP:journals/corr/abs-1903-05734}.

Even worse, several studies have highlighted the prevalence of duplication in codebases commonly used for training and evaluating source code models \cite{DBLP:journals/corr/abs-1812-06469,Lopes:2017:DMC:3152284.3133908}. Allamanis finds that the presence of duplicate examples between training and test datasets leads to accuracy metrics inflated up to 100\% \cite{DBLP:journals/corr/abs-1812-06469}. Our preprocessing step in which duplicate examples are removed from each of the corpora limits the impact of this problem.

\subsection{Performance of repetition detection}

The secondary network that detects repetition of input sequence tokens scored 0.9051 precision and 0.9071 recall on test data. 3.311\% of the keywords, identifiers, and literals in the GitHub Dart corpus are repetitions of OOV tokens occurring within the input sequence. Another 30.545\% belong to the token vocabulary and are not repetitions. This implies that repetition detection predicts a true positive that the base LM assigns zero probability in 3.003\% of examples. It will make a false positive prediction and incorrectly reassign probability mass from a non-repeat, in-vocabulary token 2.899\% of the time.

\subsection{Model size}

After an order of magnitude size reduction from post-training quantization, the token input model's size is 114M. The character input model clocks in at 65M. Both models are large but fall within an acceptable range to distribute alongside developer tools targeting programmer workstations.

\subsection{Prediction latency}

The 10,000 requests prediction benchmark was run before and after applying post-training quantization with TensorFlow Lite. Before quantization, the benchmark completed in 30 minutes and the average request time was 179 milliseconds. After quantization, the total time was reduced to 18 minutes and the average request time shortened to 109ms. 75.33\% of the requests completed in under 110ms, right around our 100ms latency target. Since prediction latency grows with vocabulary size, our benchmark results suggest that the model is as large as it can be without degrading the user experience. It is also clear from these results that similar models needing to make multiple subword predictions are too slow for code completion.

\section{Threats to validity}

As Hellendoorn et al. caution in their study of real-world code completions, the benchmarks we use to measure code completion models and their similarity to real-world usage scenarios can have significant impact on reported accuracy and usefulness in programming assistance \cite{Hellendoorn:2019:CCF:3339505.3339625}. In particular, our models were exclusively trained on already written and reviewed code. This is a common practice in software language modeling, but there is an implicit assumption that these models will generalize well to actively developed code.

Along the same lines, one positive aspect of code completion solutions driven by type analysis is that they do not suffer from \emph{concept drift}. Whereas static analysis will exhibit constant performance as new libraries, frameworks, and coding conventions evolve, it is likely that the quality of a solution based on language modeling would degrade. One example of this phenomenon from our study was a change to the Dart programming language's style guide to recommend omitting the optional \textbf{new} keyword. Although some time has passed since this change was made, our models still suggest including \textbf{new} as part of constructor invocations. The reason for this is that most of the training examples still follow the previous style recommendation. While we can address this existing problem by applying a simple transformation to our training corpora, we cannot anticipate future evolutions in our models.


\section{Conclusions}


We introduced a new approach to source code modeling that combines a character-input, token-output LM with an auxiliary network for local repetition detection and static analysis for prediction validation. Constraints on prediction latency, model size, and output validity were motivated from relevant user experience studies \cite{muaruașoiuempirical,Murphy:2006:JSD:1159169.1159396,Nielsen:1993:UE:529793,Miller:1968:RTM:1476589.1476628}.

The open vocabulary nature of source code calls for a model that can draw signal from new words at prediction time. And it is critically important that a completion model can suggest new names as programmers write new functions and libraries. We considered several prior attempts to address the open vocabulary problem in neural source code modeling \cite{DBLP:journals/corr/BhoopchandRBR16,DBLP:journals/corr/abs-1903-05734,DBLP:journals/corr/abs-1711-09573}. Networks that predict partial tokens are incompatible with the latency requirements of real-world code completion, and a token-level pointer network cannot incorporate signal from OOV tokens.

Based on these considerations, we believe that a hybrid model which learns subtoken representations but predicts whole tokens and can predict new names through repetition detection is well-suited to this task. The accuracy of such a model on large, diverse corpora of Dart code meets or exceeds all comparable source code modeling results in the literature we reviewed.

Our approach leverages a powerful combination between static analysis' ability to enumerate valid keywords and identifiers in scope and the capability from language modeling to place probability distributions over token sequences. In future work, we hope to study the modeling impact of incorporating type and program context information available through static analysis as training signal. Nguyen et al. found these elements  significant in their $\emph{n}$-gram source code model SLAMC \cite{Nguyen2013ASS}, and it's easy to imagine that these improvements could translate to neural source code models.

A large token vocabulary remains a challenge to training and prediction performance as well as model size. One of the remediations in our work was to limit the size of output vocabulary to 100k tokens, which also limits model accuracy. Although beam search over subword language model output at prediction time is a major drawback, a smaller output vocabulary of word pieces and the corresponding ability to generate identifier names not seen at training time makes such an architecture well worth exploring.

\section{Acknowledgements}

We want to thank everyone who reviewed drafts of this work and contributed valuable feedback and insightful conversation including Brian Roark, Brian Wilkerson, Izhak Shafran, Jaime Wren, Michael Fairhurst, and Phil Quitslund. Computational resources to conduct experiments as part of this research were generously provided by Google Inc.



\newpage
\bibliographystyle{ACM-Reference-Format}
\bibliography{bibfile}

\end{document}